\documentclass[prb,twocolumn,showpacs,amsmath,amssymb,superscriptaddress]{revtex4}
\usepackage[dvips]{graphicx}
\usepackage{graphics}
\usepackage{dcolumn}
\usepackage{bm}

\begin{document}
\title{Strain dependence of the Mn anisotropy in ferromagnetic semiconductors \\
observed by x-ray magnetic circular dichroism}

\author{K.~W.~Edmonds}
\affiliation{School of Physics and Astronomy, University of Nottingham, Nottingham NG7 2RD, United Kingdom}
\author{G.~van der Laan}
\affiliation{Daresbury Laboratory, Warrington WA4 4AD, United Kingdom}
\affiliation{Diamond Light Source, Chilton, Didcot OX11 0DE, United Kingdom}
\author{N.~R.~S.~Farley}
\affiliation{School of Physics and Astronomy, University of Nottingham, Nottingham NG7 2RD, United Kingdom}
\author{E.~Arenholz}
\affiliation{Advanced Light Source, Lawrence Berkeley National Laboratory, Berkeley, California 94720}
\author{R.~P.~Campion}
\affiliation{School of Physics and Astronomy, University of Nottingham, Nottingham NG7 2RD, United Kingdom}
\author{C.~T.~Foxon}
\affiliation{School of Physics and Astronomy, University of Nottingham, Nottingham NG7 2RD, United Kingdom}
\author{B.~L.~Gallagher}
\affiliation{School of Physics and Astronomy, University of Nottingham, Nottingham NG7 2RD, United Kingdom}

\date{\today}
\begin{abstract}
We demonstrate a sensitivity of the Mn $3d$ valence states to strain in the ferromagnetic semiconductors (Ga,Mn)As and (Al,Ga,Mn)As, using x-ray magnetic circular dichroism (XMCD). The spectral shape of the Mn $L_{2,3}$ XMCD is dependent on the orientation of the magnetization, and features with cubic and uniaxial dependence are distinguished. Reversing the strain reverses the sign of the uniaxial anisotropy of the Mn $L_3$ pre-peak which is ascribed to transitions from the Mn $2p$ core level to $p$-$d$ hybridized valence band hole states. With increasing carrier localization, the $L_3$ pre-peak intensity increases, indicating an increasing $3d$ character of the hybridized holes.
\end{abstract}
\pacs{75.50.Pp, 75.25.+z, 75.30.Gw, 78.70.Dm}

\maketitle

Dilute Mn impurities in III-V semiconductors such as GaAs form localized and ordered magnetic moments, while also providing polarized charge carriers. The combination of these properties results in a variety of closely correlated magnetic, transport, and structural properties. Of particular interest is the influence of epitaxial strain on the magnetic behavior. At low carrier densities (insulating regime), the magnetic easy axis lies perpendicular to the plane for compressive strain and in-plane for tensile strain, while at higher carrier densities (metallic regime) this is reversed. \cite{sawicki1} The pronounced sensitivity to strain and hole density has been described within a ${\mathbf{k}}\cdot{\mathbf{p}}$ formalism, in which the $S=\frac{5}{2}$ localized Mn moment induces an exchange splitting of the GaAs valence sub-bands. \cite{dietl} The effect of strain is to break the degeneracy of the heavy hole ($|m_j| = \frac{3}{2}$) and light hole ($|m_j| = \frac{1}{2}$) bands, giving a pronounced anisotropy of the unoccupied  $4p$ states of the host semiconductor. Calculations within this model give agreement with experimental anisotropy fields to within better than a factor of 2. However, as the tailoring of magnetic anisotropies is becoming increasingly utilized in the operation of magnetic semiconductor nanodevices, \cite{humpfner} it will be important to develop a microscopic picture of magnetic anisotropy which accounts for the full band structure of (Ga,Mn)As.

X-ray absorption (XA) spectroscopy at the Mn $L_{2,3}$ edge measures the transition probability from the Mn $2p$ core levels to $3d$ levels, thus probing directly the unoccupied valence states with Mn $3d$ character. X-ray magnetic circular dichroism (XMCD), which is the difference between XA spectra with opposite alignment of sample magnetization and x-ray helicity vector \cite{vdl1}, provides a unique method of probing magnetic moments and magnetic anisotropy on an element-specific level. This technique has brought a new level of understanding to the microscopic origin of magnetic anisotropy in thin transition metal films, \cite{weller,vdl3} and has been used previously to determine the magnetic properties of both Mn and host ions in (Ga,Mn)As. \cite{ohldag,edmonds1,keavney,jungw2} In a recent Letter, \cite{edmonds2} we identified a feature in the Mn $L_{2,3}$ XMCD spectrum from (Ga,Mn)As which shows a systematic dependence on the carrier concentration, indicating that this feature corresponds to states close to the Fermi level. Here, we demonstrate that the uniaxial anisotropy of this feature is determined by the epitaxial strain.

10-50nm thick (Ga,Mn)As and (Al,Ga,Mn)As films were grown by low temperature molecular beam epitaxy on GaAs(001) substrates. Post-growth annealing in air at 190$^{\circ}$C improves the magnetic properties of the layers through out-diffusion of interstitial defects. \cite{edmonds3} Surface Mn oxides were removed by briefly etching the films in concentrated hydrochloric acid. \cite{edmonds1} Details on film growth and characterization are given elsewhere. \cite{campion, wang} The Mn $L_{2,3}$ XA spectra were obtained on beamline 4.0.2 of the Advanced Light Source at Berkeley, CA, using 90\% circularly polarized x rays at 15~K in total electron yield mode. \cite{arenholz1} XMCD spectra were derived as difference of XA spectra measured in external fields of $\pm$0.65T.

XMCD spectra for a 10~nm thick, annealed Ga$_{0.94}$Mn$_{0.06}$As/GaAs(001) film are shown in Fig.~1. To aid comparison of spectral features, the spectra are normalized at the largest XMCD feature. As in Ref.~\onlinecite{edmonds2}, we identify two distinct anisotropies in the XMCD spectra, with cubic and uniaxial symmetry respectively. The former is evident in Fig. 1(a), where we compare spectra obtained for external magnetic fields oriented in the in-plane [110] and [100] crystalline axes, with x-rays incident at 45$^{\circ}$ to the film surface. Distinct differences in the XMCD lineshape are observed for the two field orientations, which are most visible at the peaks around 640.7~eV and 650.4~eV, and also at peak $A$ in the pre-edge region of the spectrum, shown on an expanded scale in the inset. This cubic anisotropy has been shown to be reproduced by atomic multiplet calculations for an isolated Mn $d^5$ ion in a tetrahedral environment. \cite{edmonds2} Since the multiplet calculations take no account of the (Ga,Mn)As band structure, the cubic anisotropy is a generic effect for localized transition metal ions in a crystalline host.

\begin{table*}
\caption{\label{tab:table1} Summary of the structural, magnetic, and electronic sample properties.}
\begin{ruledtabular}
\begin{tabular}{llccc}
  Sample & Composition & Strain & Easy magnetization direction & Transport properties \\
  \hline
  (i)  & Ga$_{0.98}$Mn$_{0.02}$As/GaAs(001) & compressive & in-plane & metallic \\
  (ii) & Ga$_{0.92}$Mn$_{0.08}$As/GaAs(001) & compressive & in-plane & metallic \\
  (iii)& Ga$_{0.74}$Al$_{0.2}$Mn$_{0.06}$As/GaAs(001) & compressive & out-of-plane & insulating \\
  (iv) & Ga$_{0.95}$Mn$_{0.05}$As/(In,Ga)As/GaAs(001) & tensile & out-of-plane & metallic \\
\end{tabular}
\end{ruledtabular}
\end{table*}

\begin{figure}[h]
\centering
\includegraphics[trim = 50mm 40mm 50mm 45mm,clip, width=80mm, angle=0]{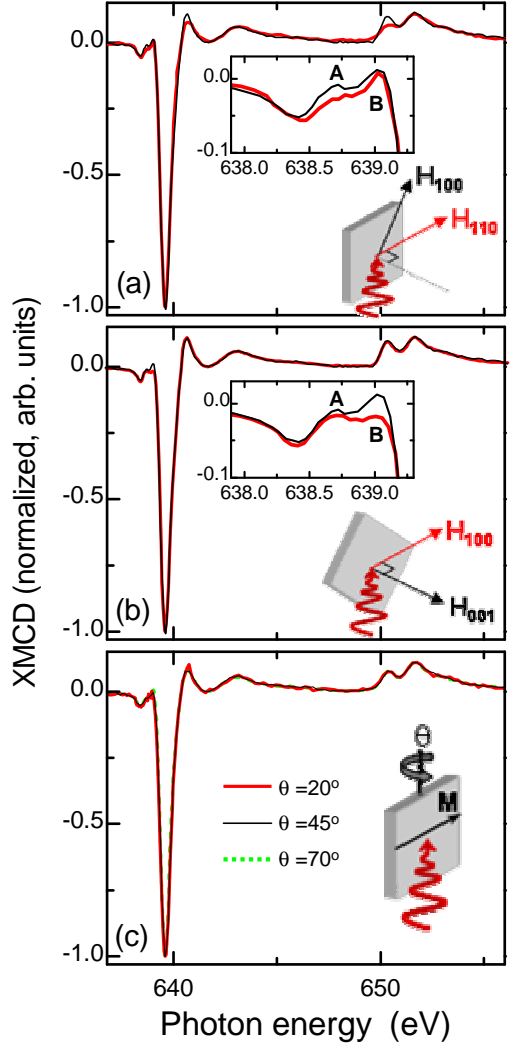}
\caption{\label{fig:Fig1} (Color online) Mn $L_{2,3}$ XMCD spectra of Ga$_{0.94}$Mn$_{0.06}$As on GaAs(001)
obtained in three distinct experimental geometries: (a) the x-rays are incident at 45$^{\circ}$ with respect to
the film surface and the external magnetic field is applied in the plane of the film, either along the [100]
direction (thin black line) or the [110] direction (thick red/grey line); (b) the x-rays are incident at
45$^{\circ}$ to the film surface and the external field applied in either the in-plane [100] direction (thin
black line) or the out-of-plane [100] direction (thick red/grey line); (c) The XMCD is measured at remanence
with the magnetization in the plane of the film along the [110] axis, for x-ray incidence angle $\theta
=20^{\circ}$ (thick red/grey line), 45$^{\circ}$ (thin black line), and 70$^{\circ}$ (broken green line). The
data are normalized to a peak $L_3$ XMCD intensity of $-1$. The pre-edge region is expanded in the upper insets
to (a) and (b). The lower insets illustrate the experimental geometries. }
\end{figure}

\begin{figure}[h]
\centering
\includegraphics[trim = 60mm 70mm 60mm 65mm,clip, width=80mm, angle=0]{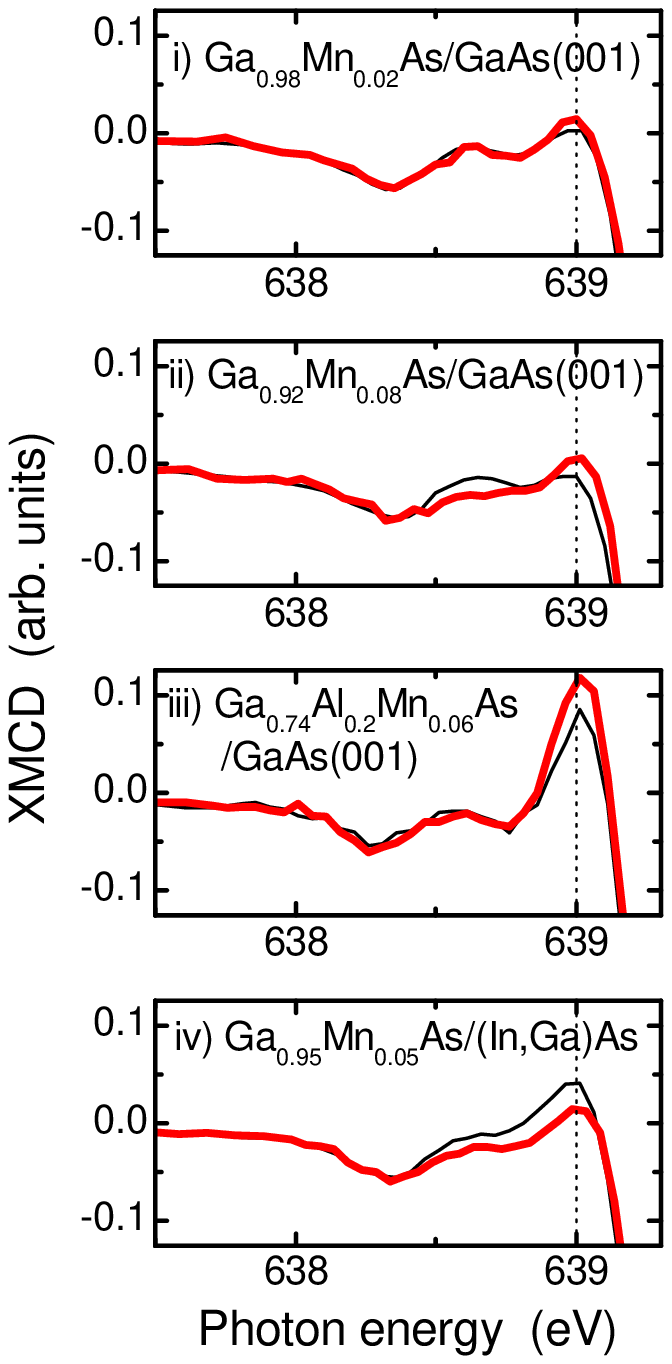}
\caption{\label{fig:Fig2} (Color online) Mn $L_3$ pre-edge XMCD spectra for (i) Ga$_{0.98}$Mn$_{0.02}$As on
GaAs(001); (ii) Ga$_{0.92}$Mn$_{0.08}$As on GaAs(001); (iii) Ga$_{0.74}$Al$_{0.2}$Mn$_{0.06}$As on GaAs(001);
(iv) Ga$_{0.95}$Mn$_{0.05}$As on (In,Ga)As. Measurements are with external field parallel to the x-ray incidence
direction and at normal incidence ($\theta =90^{\circ}$, thin black line) or grazing incidence ($\theta
=25^{\circ}$, thick red/grey line). }
\end{figure}

The uniaxial anisotropy is evident in Fig. 1(b), where spectra are compared for magnetic fields oriented along the in-plane and out-of-plane $\langle$100$\rangle$ axes. This anisotropy is most pronounced at the pre-edge peak $B$, at around 0.4~eV below the main $L_3$ resonance. The peak $B$ is found to depend on the angle between the magnetization direction and the sample plane, independent of the crystalline axes, and is absent in the calculated spectrum of the $d^5$ multiplet. \cite{edmonds2} The pre-peak is observed for all (Ga,Mn)As samples studied, but is largest for samples on the insulating side of the metal-insulator transition. The origin of this feature will be discussed below.

In Fig.~1(c), the XMCD spectra are recorded at remanence after applying a magnetic field along the in-plane [110] crystalline axis (the easy magnetic axis in annealed (Ga,Mn)As \cite{sawicki2}) for different angles between the photon incidence direction and the [110] direction. In this case, no significant differences are observed between the spectra. Therefore, the XMCD line shape is dependent on the alignment of the magnetization with respect to its crystalline axis, but is independent of the photon incidence angle if the magnetization orientation is fixed.

We investigate the dependence of pre-peak $B$ on the sample parameters by comparing XMCD spectra from four (Ga,Mn)As films: (i) Ga$_{0.98}$Mn$_{0.02}$As on GaAs(001), (ii) Ga$_{0.92}$Mn$_{0.08}$As on GaAs(001), (iii) Ga$_{0.74}$Al$_{0.2}$Mn$_{0.06}$As on GaAs(001), and (iv) Ga$_{0.95}$Mn$_{0.05}$As on a (In,Ga)As buffer layer on GaAs(001). Films (i) to (iii) are under compressive strain, which increases approximately linearly with Mn content. \cite{shen,zhao} For film (iv), growth on the larger lattice constant (In,Ga)As results in tensile strain. This film has a perpendicular-to-plane easy magnetic axis, as is typically observed for tensile-strained (Ga,Mn)As. \cite{sawicki1,wang,shen} For films (i) and (ii), the easy axis is in-plane, as is typical for compressive-strained (Ga,Mn)As. For film (iii), the effect of Al incorporation is to shift the valence band maximum to lower energies, resulting in an stronger binding of the hole to the Mn ion \cite{masek} and a transition from metallic to insulating behavior. This film has an out-of-plane easy magnetic axis, as has been observed previously for (Ga,Al,Mn)As \cite{takamura} as well as insulating low-hole-density (Ga,Mn)As under compressive strain. \cite{sawicki1} The sample properties are summarized in Table~I.

Figure~2 shows the Mn $L_3$ pre-peak structure for the four films, obtained with the external field collinear with the x-ray beam in normal incidence ($\theta$ = 90$^\circ$) or grazing incidence ($\theta$ = 25$^\circ$), where $\theta$ is defined in the inset to Fig.~1(c). For the uniaxial anisotropy of prepeak $B$ a striking dependence on the magnitude and direction of the epitaxial strain is observed. For the compressive strained films (i), (ii), and (iii), the pre-edge peak at 639~eV is largest when the magnetization is close to in-plane, while for the film under tensile strain (iv), the peak is largest when the magnetization lies perpendicular to the plane. Furthermore, the anisotropy in the pre-peak is smallest for the film with 2\% Mn, which has the smallest strain. In contrast, the spectral features showing cubic anisotropy such as peak $A$ are similar for all four samples, although the size of the cubic anisotropy varies.

The magnetic anisotropy of III-V magnetic semiconductors depends sensitively on the interplay of strain and hole density. \cite{sawicki1} Therefore, in spite of the opposite strain direction in films (iii) and (iv), the easy magnetization axis is out of plane in both samples due to the very different hole density.  Magnetic anisotropy leads to an anisotropy in the magnetic dichroism spectra, since both are related to the anisotropic part of the orbital moment. \cite{weller,vdl3} However, comparison of the XMCD spectra in Fig.~2 shows that the pre-peak anisotropy is not directly related to the magnetic anisotropy but is determined only by strain, being opposite in sign for films (iii) and (iv).

In $L_{2,3}$ absorption, electric-dipole transitions excite an electron from a $p$ core level to unoccupied states with $s$ and $d$ character. The strain-dependent uniaxial anisotropy observed in the XMCD may arise from the anisotropy of the magnetic moment of these states. Since $s$ states are isotropic, only the $d$ states can give anisotropy in the XMCD. Additionally, a Mn $3d^5$ state with $S=\frac{5}{2}$ is not expected to show a uniaxial strain splitting due to the absence of an orbital moment. Therefore, the strain-dependence of the XMCD is attributed to deviations from a pure $3d^5$ configuration, resulting from hybridization with neighboring As $4p$ orbitals.

In tetrahedral symmetry, the Mn $t_2$ and $e$ orbitals are separated in energy by the crystal field interaction. The $t_2$ states hybridize with the As $s$, $p$ orbitals, while the $e$ orbitals are non-bonding. Strain imposes a further symmetry breaking, splitting the $d_{xy}$ orbitals from the $d_{zx}$ and $d_{yz}$ orbitals in the $t_2$ state. However, this splitting cannot by itself give rise to the distinct uniaxial-anisotropy of the pre-peak, since within a single-ion model all parts of the multiplet structure should display the same symmetry \cite{vdl2}. In contrast, features with cubic and uniaxial symmetry can be clearly distinguished in the Mn $L_{2,3}$ XMCD. These features must therefore correspond to Mn $3d$ states with fundamentally different characters.

We ascribe the cubic features of the spectrum to the localized Mn states, which lie deep in the valence band, and peak $B$ to a $p$-$d$ hybridized band-like feature, which intersects with the Fermi level $E_F$. Such a band emerges in {\it ab-initio} calculations of the (Ga,Mn)As local density of states, which predict a small contribution of Mn $d$ levels to the density of states at $E_F$. \cite{sandratskii,schulthess} Further evidence for this interpretation comes from angle-resolved photoemission studies in which a dispersionless Mn-induced band near $E_F$ was reported. \cite{okabayashi} The ground state of the Mn ion can be written as a mixed state $\alpha|3d^4\rangle + \beta|3d^5\underline{h}\rangle + \gamma|3d^6\underline{h}^2 \rangle$, where $\alpha$, $\beta$, and $\gamma$ are wavefunction coefficients, and $\underline{h}$ represents combination of appropriate symmetry of states characterizing a $p$-$d$ hybridized valence band hole. \cite{okabayashi} Dipole transitions are allowed to a mixed final state $\alpha'|\underline{2p}3d^5 \rangle + \beta'| \underline{2p}3d^6\underline{h} \rangle + \gamma'| \underline{2p}3d^7\underline{h}^2 \rangle$, where $\underline{2p}$ denotes the core hole.  The strong core-valence interaction lowers the energy of the final state, so that it is pulled below the main $3d$ band width and obtains a localized character.

For transitions of the type $3d^n\underline{h} \to \underline{2p}3d^{n+1}\underline{h}$, the $2p$ electron is excited into an empty $3d$ state with crystal field symmetry. In this transition the $\underline{h}$ state is a spectator, and the $\underline{2p}3d^{n+1}$ final state exhibits a multiplet structure.  For (Ga,Mn)As, the ground state has predominantly $ 3d^5\underline{h}$ character, corresponding to the configuration of the lowest energy, which gives rise to the $3d^5$ multiplet structure in the XMCD spectrum. The pre-peak, on the other hand, can be ascribed to transitions of the type $3d^n\underline{h} \to \underline{2p}3d^n$, where the $2p$ electron is excited into an empty state in the $p$-$d$ hybridized (Ga,Mn)As valence band. The final state is localized by the strong Coulomb interaction with the $2p$ hole, leading to a loss of its band character. It therefore has the uniaxial symmetry of the (Ga,Mn)As valence band at $k=0$.

The XMCD signal is proportional to the total magnetic moment of the unoccupied state selected by the excitation.
\cite{vdl1} The pre-peak has opposite sign from the main peak since it originates from states near $E_F$ with opposite polarization to the Mn local moment. Also, the pre-peak is largest when the magnetization is perpendicular-to-plane (in-plane) for tensile (compressive) strain. This is opposite to the anisotropy at the valence band maximum predicted by ${\mathbf{k}}\cdot{\mathbf{p}}$ theory, \cite{dietl} because the charge quadrupole moment giving rise to strain-splitting is opposite in the Mn $t_2$ states compared to the As $p$ states for magnetization along the same direction. Finally, the pre-edge peak is largest for insulating ferromagnetic films such as film (iii), due to localization of the hybridized holes around their parent Mn ions resulting in an increasing Mn $3d$ (as opposed to As 4$p$) character. Therefore, the cross-section for $3d^n\underline{h} \to \underline{2p}3d^n$ transitions will increase.

In summary, we performed a systematic investigation on the strain dependence of the uniaxial anisotropy observed in XMCD for dilute Mn impurities in III-V semiconductors. The anisotropy is positive for compressive strain and negative for tensile strain, irrespective of the direction of the magnetic easy axis in the film. The spectral feature with uniaxial anisotropy is ascribed to transitions into a hybridization of Mn $3d$ and host valence band states, which appears in the calculated (Ga,Mn)As density of states and is reported in angle-resolved photoemission measurements. \cite{sandratskii, schulthess, okabayashi} The observation of such a feature with XMCD demonstrates that this state is fully participating in the ferromagnetism, and attests to its Mn $3d$ character. These findings provide new insight into the electronic structure of ferromagnetic semiconductors, and the interplay of strain and magnetization in such systems.

The Advanced Light Source is supported by DOE. Funding from the UK EPSRC (EP/C526546/1), STFC, and the Royal
Society is acknowledged.

\end{document}